\documentclass[journal]{IEEEtran}
\usepackage{cite}
\usepackage{color}
\makeatletter

\newcommand{\Rmnum}[1]{\expandafter\@slowromancap\romannumeral #1@}
\makeatother
\usepackage{graphicx}
\usepackage{lipsum,multicol}
\usepackage{amsmath}

\usepackage{mathrsfs}
\usepackage{amssymb}
\usepackage[cmintegrals]{newtxmath}
\usepackage{algorithm}
\usepackage{algorithmic}

\usepackage{array}
\usepackage{booktabs}
\usepackage{tabularx}
\usepackage{multirow}
\usepackage[pscoord]{eso-pic}
\usepackage[numbers,sort&compress]{natbib}
\ifCLASSINFOpdf
\else
\fi
\begin{document}

\title{Multilayer Routing and Resource Assignment in Spatial Channel Networks (SCNs): Oriented Toward the Massive SDM Era}

\author{Mingcong~Yang~\IEEEmembership{Member,~OSA}, Qian~Wu, Maiko Shigeno, Yongbing~Zhang 
\thanks{This work has been submitted to the IEEE for possible publication. Copyright may be transferred without notice, after which this version may no longer be accessible.}}	

\maketitle


\begin{abstract}
In the past few decades, optical transport networks (OTNs) have undergone significant evolution, from the earliest wavelength-division multiplexing (WDM) OTNs to elastic optical networks (EONs) and later to space-division multiplexing (SDM) OTNs, to address the continuous growth of Internet traffic. By 2024, Pbps-level OTNs are expected, far exceeding the capacity limit of single-mode fibers. The massive SDM era is on the horizon. In this context, newly designed OTNs called spatial channel networks (SCNs), which achieve high cost efficiency by means of practical hierarchical optical cross-connects, have recently been proposed. However, the evolution of OTNs will simultaneously present challenges related to resource allocation in networking. For instance, with the evolution from WDM-OTNs to EONs, the resource allocation problem was transformed from the routing and wavelength assignment (RWA) problem to the routing and spectrum assignment (RSA) problem due to the additionally introduced constraint of spectrum contiguity. Similarly, specially designed algorithms are also expected to be essential for addressing the resource allocation problem in SCNs. In this paper, we define this new problem as the routing, spatial channel, and spectrum assignment (RSCSA) problem. We propose an integer linear programming (ILP) model and a heuristic algorithm to solve the RSCSA problem. We examine the performance of the proposed approaches via simulation experiments. The results show that both proposed approaches are effective in finding the optimal solutions or solutions close to the lower bounds. To the best of our knowledge, this is the first work to focus on the problem of resource allocation in SCNs.
\end{abstract}

\section{Introduction}
With the development and increasing popularity of cloud computing, video-on-demand (VoD), the Internet of Things (IoT), and other emerging Internet services, network traffic is growing at an extremely rapid rate \cite{winzer2017scaling}. Consequently, since the deployment of wavelength-division multiplexing (WDM) optical transport networks (OTNs) at the beginning of the 2000s, OTNs have undergone several significant evolutions to support the rapid increase in network traffic. The first evolution, from traditional WDM-OTNs to elastic optical networks (EONs), which was induced by the introduction of advanced technologies such as orthogonal frequency-division multiplexing, Nyquist wavelength-division multiplexing, and distance-adaptive modulation, greatly increased the spectrum efficiency \cite{gerstel2012elastic}. Moreover, to overcome the capacity limit of conventional single-mode fibers (SMFs), that is, the so-called nonlinear Shannon limit, space-division multiplexing (SDM) technology was later proposed, which motivated the further evolution of OTNs from EONs to SDM-OTNs \cite{winzer2013spatial}.

From the networking perspective, the resource allocation problem also changed several times in accordance with the new features introduced by the evolution of OTNs. In WDM-OTNs, the resource allocation problem is called the routing and wavelength assignment (RWA) problem \cite{ramaswami1995routing}. A lightpath, composed of a routing path and a wavelength, must be assigned to each connection request. The assigned wavelength must be consistent along the entire lightpath (unless wavelength conversion is allowed) and be nonoverlapping with the other wavelengths on each fiber link. These constraints are the so-called wavelength continuity and wavelength nonoverlap constraints, respectively.

Compared with WDM-OTNs, more flexible spectrum divisions are possible in EONs, such as the 12.5 GHz frequency slices (FSs) that conform to the G.694.1 standard recommended by the International Telecommunication Union Telecommunication Standardization Sector (ITU-T) \cite{ITU-T}. In combination with the application of bandwidth-variable transceivers (BVTs) \cite{sambo2015next} and optical cross-connects (BV-OXCs) \cite{kozicki2010optical} in EONs, it is possible to satisfy connection requests with varying bit rates and to flexibly establish lightpaths by using different numbers of FSs as needed, thus achieving higher spectrum efficiency. However, a disadvantageous consequence of this approach is that the FSs assigned to each connection request should be contiguous, which introduces an additional constraint of spectrum contiguity. Therefore, the resource allocation problem is transformed into the routing and spectrum assignment (RSA) problem in EONs \cite{christodoulopoulos2011elastic, shen2016survivable}.

Although EONs are promising OTNs that achieve more efficient utilization of spectrum resources compared to traditional WDM-OTNs, the growth in the transmission capacity of standard SMFs has dramatically slowed because the current transmission capacity per fiber is approaching the nonlinear Shannon limit of the existing SMFs. Nevertheless, the volume of Internet traffic is expected to continue to strongly increase in the future, inexorably reaching this capacity limit \cite{essiambre2010capacity}. Thus, as a viable solution for overcoming this limit, SDM technology has emerged, the basic concept of which is to expand the available space lanes (SLs) from the current single SL (i.e., an SMF) to multiple parallel SLs to increase the available spectrum resources \cite{richardson2013space}. This expansion will enable us to assign spectrum resources straddling both the spectral and spatial domains, which will again make the resource allocation problem more complicated because the appropriate SL(s) should be assigned to each lightpath simultaneously with the assignment of the routing path and spectrum. Therefore, in this case, the resource allocation problem becomes the routing, spectrum, and space assignment (RSSA) problem \cite{klinkowski2018survey}.

As reported in Ref.~\cite{winzer2017scaling}, the compound annual growth rates (CAGRs) of the aggregate router blade interface rate have been approximately 40\% in recent years. By 2024, the implementation of an optical interface rate of up to 10 Tbps is expected to be required. Moreover, considering that the interconnections between adjacent nodes are expected to consist of dozens or even hundreds of SLs (fibers/cores) in the near future, Pbps-level OTNs are anticipated. The massive SDM era is on the horizon. However, considering that the total bandwidth of the C-band is approximately 4 THz per fiber/core for SMFs or multicore fibers (MCFs), for ultralong-haul optical transmission, the enormous bandwidth requirement of a 10 Tbps connection request will exceed the entire C-band for such fibers \cite{jinno2019spatial}. Wavelength switching will no longer be necessary to transmit such connection requests because the entire fiber/core will become a logical end-to-end interface to serve an ultrahigh-capacity optical data stream, which will be routed as a single entity by optical bypass technology. In this context, spatial channel networks (SCNs) \cite{asano2018cost, jinno2018spatial, jinno2019spatial, jinno2019required, jinno2019opportunities, jinno2019spatial2} have been proposed as an economical and realistic solution oriented toward the future massive SDM era. Nevertheless, similar to what has happened heretofore, the evolution of OTNs will present further opportunities for addressing the resource allocation problem in networking and will also pose corresponding challenges. Dedicated algorithms considering the features of the newly designed OTNs will be essential to address the corresponding resource allocation problem \cite{jinno2019spatial}.

In this paper, reviewing the significant evolution of OTNs that has occurred over the past few decades, we focus on the static resource allocation problem in SCNs, which we define as the routing, spatial channel, and spectrum assignment (RSCSA) problem. In Section II, we identify the novel features of SCNs from the networking perspective. In Section III, we define the RSCSA problem and prove that it is NP-hard. We also clarify the constraints corresponding to the features of SCNs in detail. In Sections IV and V, we propose an integer linear programming (ILP) model and a heuristic algorithm, respectively, for solving the RSCSA problem. In Section VI, we evaluate the performance of the two proposed approaches via simulation experiments. Finally, in Section VII, we conclude the paper and present prospects for future work.

\section{Spatial channel networks}
\begin{figure*}[t]
	\begin{center}
		\includegraphics[width=16cm]{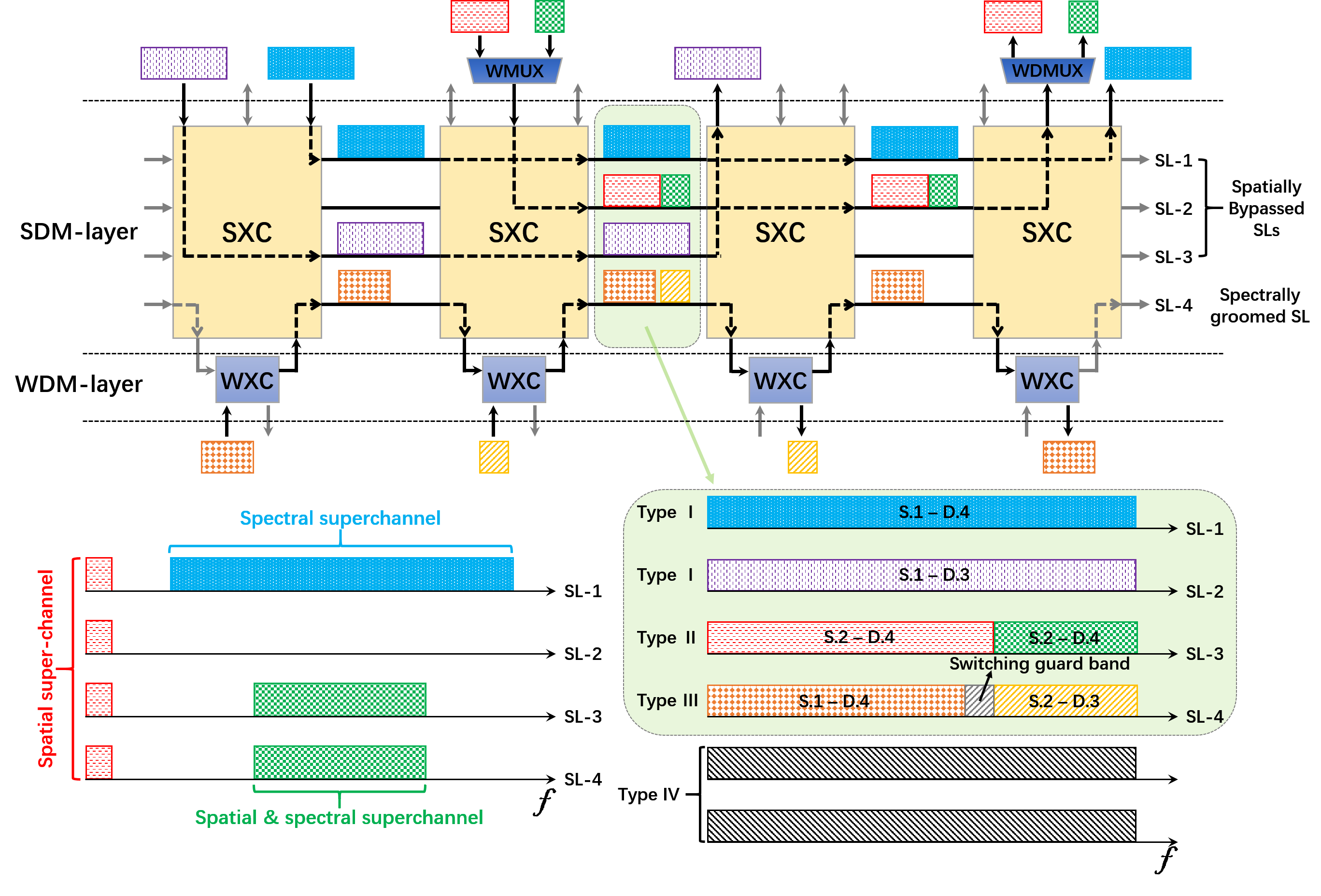}
	\end{center}
	\caption{Illustration of the spectral and spatial SpChs in SDM-OTNs vs. the SChs in SCNs.}
	\label{fig-SCh}
\end{figure*}
As stated in Section I, network traffic has grown at an extremely rapid rate over the past few decades, which has inevitably compelled the development of optical transmission technologies, as well. As shown in the bottom left of Fig.~\ref{fig-SCh}, spectral superchannel transmission technology, which comprises several adjacent optical carriers (OCs) without switching guard bands (SW-GBs) between them, has been effectively applied in EONs, leading to higher spectrum efficiency \cite{bosco2010performance}. In addition, the expansion of the SLs in SDM-OTNs enables us to allocate OCs that span multiple SLs but share a single laser source to create a spatial superchannel, leading to higher cost efficiency. Of course, any suitable hybrid combination of the two types of superchannels above, a so-called spectral and spatial superchannel, as shown in Fig. \ref{fig-SCh}, is also feasible for use in SDM-OTNs \cite{yang2019evaluation}.

Considering the aforementioned 10 Tbps client interface rate that is anticipated to be achieved by 2024, one hundred 32 Gbaud DP-QPSK OCs (each supporting 100 Gbps) will be required to establish such a connection request for long-haul transmission, or other combinations may be suitable for a shorter distance, such as twenty-five 64 Gbaud DP-16-QAM OCs \cite{winzer2017scaling, jinno2019spatial}. We can see that a total spectrum of 3.2 THz is required in the ideal case (i.e., with the ideal Nyquist shaping and a gridless spectrum) for a 10 Tbps DP-QPSK spectral superchannel, and the entire C-band can accommodate only one such superchannel. This indicates that wavelength switching support will no longer be necessary for every SL, since after a few more years, the spectral superchannel used to serve a single connection request may require the entire C-band spectrum. SCNs with hierarchical optical cross-connects (HOXCs) have therefore been recently proposed \cite{asano2018cost, jinno2018spatial, jinno2019spatial, jinno2019required, jinno2019opportunities, jinno2019spatial2}.

\subsection{Spatial channels}
First, we introduce the concept of spatial channels (SChs) in SCNs. An SCh is defined as an ultrahigh-capacity optical data stream that occupies a large amount of spectrum, and it can be optically routed in an end-to-end manner as a single entity through spatial cross-connects (SXCs) (called spatial bypass in SCNs) \cite{asano2018cost, jinno2018spatial, jinno2019spatial, jinno2019required, jinno2019opportunities, jinno2019spatial2}. It should be noted that the concepts of SChs and superchannels are different, although in some previous works, the abbreviation SCh has also been used for superchannels. In this paper, SpCh is used as the abbreviation for the term `superchannel' to avoid confusion. As shown in Fig.~\ref{fig-SCh}, there are four types of SChs, which are listed as follows:
\begin{itemize}
\item Type I: An SCh that carries a single high-capacity spectral SpCh (shown in blue and purple in Fig.~\ref{fig-SCh}). SChs of Type I can be routed in an end-to-end manner through spatial bypass without wavelength switching.

\item Type II: An SCh that carries multiple spectral SpChs established between the same source-destination pair (shown in red and green). SChs of Type II can also be end-to-end spatially bypassed, while multiple spectral SpChs belonging to such an SCh can be allocated without SW-GBs.

\item Type III: An SCh that carries multiple spectral SpChs established between different source-destination pairs (shown in orange and yellow). These spectral SpChs are added/dropped by the wavelength cross-connects (WXCs) at intermediate node(s), and thus, SW-GBs are required between them.

\item Type IV: An SCh that carries a single ultrahigh-capacity spatial and spectral SpCh (shown in black), which occupies multiple SLs. However, in this paper, we do not consider SChs of Type IV because such an SCh can be equivalently treated as multiple SChs of Type I, each of which can be routed independently.
\end{itemize}

\subsection{Hierarchical optical cross-connects}
\begin{figure*}[t]
	\begin{center}
		\includegraphics[width=17cm]{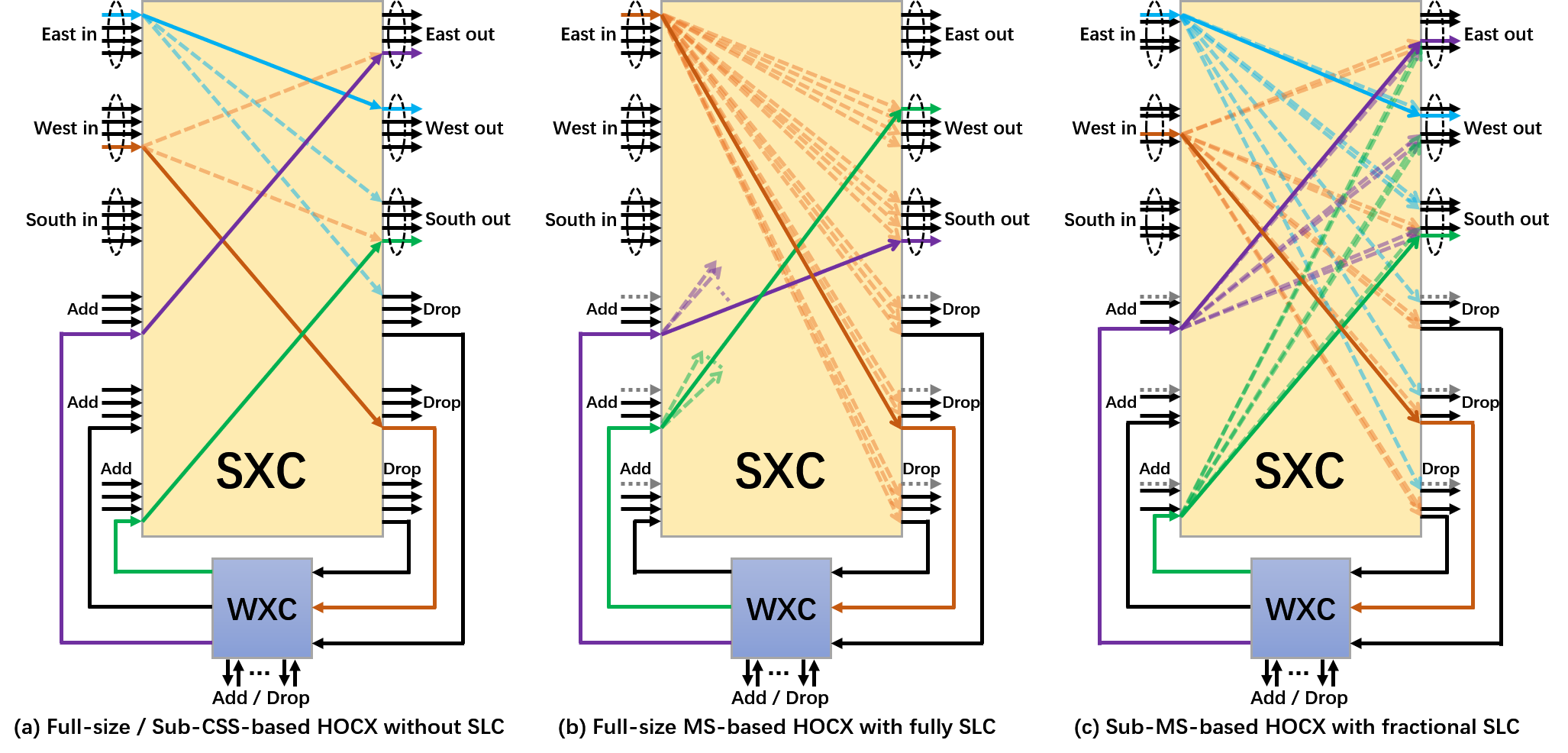}
	\end{center}
	\caption{Illustration of four HOXCs proposed for use in SCNs from the networking perspective. Solid arrow: active switching; dotted arrow: possible switching.}
	\label{fig-SCN}
\end{figure*}
As shown in Fig.~\ref{fig-SCh}, in an SCN, the switching layer is divided into an SDM layer and a WDM layer to achieve higher cost efficiency. SChs of Type I, Type II, and Type IV are spatially bypassed without passing through the WDM layer. In Ref.~\cite{jinno2019spatial}, four different types of HOXCs, which support different degrees of cost efficiency, routing flexibility, and scalability, have been proposed to achieve this functionality. In fact, the concept of HOXCs was first proposed in the late 1990s \cite{harada1999hierarchical, saleh1999architectural}, and some efforts were made in this direction before the concept of SDM-OTNs began to gain in popularity \cite{cao2006framework, ishii2009an}. This paper aims to identify the distinctive features of SCNs from the networking perspective but does not focus on explaining the detailed architectures of HOXCs for SCNs or comparing them with previous architectures. Readers can refer to Ref.~\cite{jinno2019spatial} for more detailed related information.

Fig.~\ref{fig-SCN} illustrates the HOXCs proposed for use in SCNs, which are implemented on the basis of full-size core-selective switches (CSSs) \cite{jinno2019architecture}, sub-CSSs, full-size matrix switches (MSs) \cite{sohma2006silica}, and sub-MSs.

\begin{itemize}
\item Full-size CSS-based HOXC: The full-size CSS-based HOXC is the most cost-efficient solution among the four HOXCs. It also supports the scaling up of the nodal degree. However, space lane change (SLC) is not supported by this HOXC. For example, as shown in Fig.~\ref{fig-SCN}.(a), if we assume that the logical indices of the SLs (fibers/cores) are the same on each link, then an SCh that enters an intermediate node can be switched only to output ports (including drop ports) with the same index.

\item Sub-CSS-based HOXC: The sub-CSS-based HOXC is also a cost-efficient solution but costs more than the full-size CSS-based HOXC. However, it supports the scaling up of not only the nodal degree but also the number of SLs per degree in compensation for its additional cost. In addition, it has the same features as the full-size CSS-based HOXC from the networking perspective, as shown in Fig.~\ref{fig-SCN}.(a).

\item Full-size MS-based HOXC: The full-size MS-based HOXC is the solution that provides the highest routing flexibility among the four HOXCs. As shown in Fig.~\ref{fig-SCN}.(b), this HOXC allows an SCh to be switched to any output port (including drop ports). However, it is also the costliest solution and does not support the scalability of the nodal degree and the number of SLs (per degree). It is worth noting that since the full-size MS-based HOXC supports full SLC, a single add/drop port can be used to add/drop SChs to/from SLs with different indices (at different time points). Therefore, the add/drop port counts can be reduced to some extent (an example is illustrated by the gray dotted arrow).

\item Sub-MS-based HOXC: The sub MS-based HOXC is a compromise solution relative to the full-size MS-based HOXC. In this case, the SLs are divided into multiple groups (e.g., two groups in the example shown in Fig.~\ref{fig-SCN}.(c)), and SLC is available within each group. Compared to the full-size MS-based HOXC, this solution sacrifices some routing flexibility in exchange for support for the scalability of the number of SLs per degree and a considerable cost savings. Nevertheless, it is still much costlier than either of the two CSS-based HOXCs.
\end{itemize}

In summary, the above four HOXCs show various differences in cost efficiency, routing flexibility, and scalability. However, all of them cost less than conventional OXCs, which require wavelength switching support on each SL in SDM-OTNs. In this paper, we consider only SCNs implemented on the basis of full-size/sub-CSS-based HOXCs (as shown in Fig.~\ref{fig-SCN}.(a)) and defer the consideration of applications of the two MS-based HOXCs to future research. This is because the two CSS-based HOXCs offer significantly higher cost efficiency -- readers can refer to the cost assessments in Refs.~\cite{asano2018cost} and \cite{jinno2019spatial} for more details -- and scalability than the two MS-based HOXCs do and thus are considered more suitable for use in future commercial SCNs.

\section{Routing, spatial channel, and spectrum assignment (RSCSA) problem}
\subsection{Introduction to the RSCSA problem}
Similar to the RWA problem in WDM-OTNs, the RSA problem in EONs, and the RSSA problem in SDM-OTNs, the RSCSA problem can be subdivided into two main cases: the dynamic case and the static case.

In the dynamic case, which emerges during network operation, it is assumed that the connection requests are unknown in advance and that they stochastically arrive and disappear one by one. The resources required to serve connection requests are assigned dynamically in accordance with the current state of the network. The objective of the dynamic RSCSA problem is to minimize the network blocking probability (BP) or to maximize the network throughput while maintaining an acceptable BP (e.g., 1\%) \cite{rumipamba2018on, rumipamba2018space}, which is the same as the objectives of the previous dynamic RWA, RSA, and RSSA problems.

In the static case, which mainly relates to the network planning phase, a traffic matrix that contains a set of connection requests that must be served in the network is known in advance, and resources must be assigned to all of these connection requests simultaneously. In the static RSCSA problem, the main objective is to minimize the number of SLs that are used/required in the network, for the following three reasons:

\begin{itemize}
\item Minimizing the number of FSs that are used/required (or the maximum index of these FSs) in the network, as is done in the static RWA, RSA, and RSSA problems, is pointless in this case because in an SCN, each connection request is transmitted by an SCh, which may occupy the entire C-band spectrum.

\item Minimizing the number of SLs used is equivalent to maximizing the number of SLs in the network that are not occupied and thus are available for future connection requests -- assuming that the network scenario is semidynamic, we optimize the network by reassigning the currently established connections as a `static' set, and any connection requests that subsequently arrive in the network are handled dynamically \cite{zhang2007optimization}. Therefore, minimizing the number of SLs used reduces the level of congestion in the network.

\item The last reason is that there are many different possible types of SCN systems. Note that scalability of the SLs is not supported by all types of HOXCs, and in general, a system with 20 SLs is much cheaper than one with 40 SLs. Therefore, if we can reduce the number of SLs required to below 20 during the network planning phase, great cost savings can be achieved.
\end{itemize}

Another objective of the RSCSA problem, although with a lower priority, is to minimize the number of SLs with wavelength switching support that are used/required, for the following two reasons:

\begin{itemize}
\item As stated before, compared to SDM-OTNs, the key feature of SCNs is that wavelength switching support is not necessary on every SL because some connection requests can be transmitted by SChs of Type I and Type II, which can be spatially bypassed at intermediate nodes. As shown in Fig.~\ref{fig-SCN}, the number of SLs with wavelength switching support has a one-to-one relationship with the number of deployed WXCs. Therefore, during the network planning phase, minimizing the number of required SLs with wavelength switching support is equivalent to minimizing the number of required WXCs at HOXCs. Between two HOXCs that support the same number of SLs, the one with fewer WXCs will certainly cost less.

\item As introduced in Section II-A, there are four types of SChs, and from the networking perspective, an SCh of Type IV can be treated as multiple independent SChs of Type I. SChs of Type I and Type II can be spatially bypassed at intermediate nodes using SLs without wavelength switching support. However, if the available SLs without wavelength switching support are inadequate, SLs with wavelength switching support can also be used. In contrast, SChs of Type III can pass only through SLs with wavelength switching support. Therefore, in the semidynamic scenario, for two solutions to the RSCSA problem that require an equal number of SLs, the one that uses fewer SLs that support wavelength switching is preferred -- the more idle SLs with wavelength switching support there are, the higher the possibility of satisfying more subsequent connection requests.
\end{itemize}

In summary, the static RSCSA problem is a multiobjective problem in which the decision on how to allocate resources, such as routing paths, SLs, modulation formats, and spectrum, for each connection request should be jointly made in an offline manner.

\subsection{NP-hardness of the RSCSA problem}
In this subsection, we prove the NP-hardness of the RSCSA problem by reducing the RWA problem for traditional WDM-OTNs to the related RSCSA problem.

\begin{figure}[!htbp]
	\begin{center}
		\includegraphics[width=7cm]{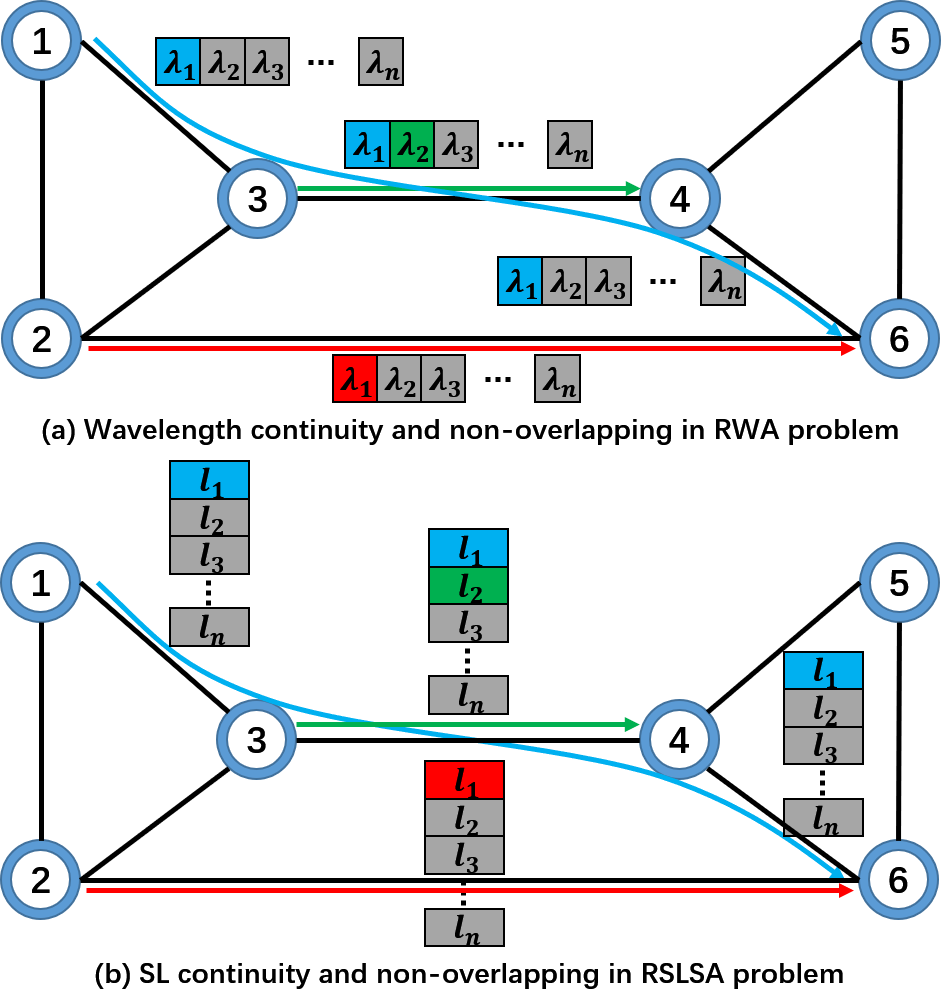}
	\end{center}
	\caption{Comparison between the RWA problem and the RSCSA problem.}
	\label{fig-NP-RSCSA}
\end{figure}

The RWA problem is a well-known NP-hard problem \cite{wauters1996design}. An instance of the RWA problem includes a set of connection requests $r \in R$ and a set of wavelengths $\lambda \in \Lambda$. The objective is to assign a routing path $p_r$ and a wavelength $\lambda_r$ to each $r \in R$ while minimizing the number of wavelengths that are used/required in the network ($\lambda_{\text{\it max}}$). In addition, the assignments should comply with the wavelength continuity and nonoverlap constraints. To solve the RWA problem in a form that is equivalent to the RSCSA problem, we consider the network scenario shown in Fig.~\ref{fig-NP-RSCSA}.

We assume that a set of connection requests $r \in R$ and a set of SLs $l \in L$ ($L \rightarrow \Lambda$) are given. Note that the two CSS-based HOXCs considered in this paper do not support SLC. Therefore, the wavelength continuity constraint is converted into an SL continuity constraint in the RSCSA problem, and a corresponding SL nonoverlap constraint should also be satisfied. Here, we simplify the RSCSA problem by ignoring the second (minor) objective and assume that each $r \in R$ exactly occupies the entire C-band of a single SL. Thus, we should assign a routing path $p_{r}$ and an SL $l_r$ ($l_r \rightarrow \lambda_r$) to each $r \in R$ while minimizing the number of SLs that are used/required in the SCN ($l_{\text{\it max}}$; $l_{\text{\it max}} \rightarrow \lambda_{\text{\it max}}$). In this case, if we were able to optimally solve the simplified RSCSA problem, we would also obtain the optimal solution to the RWA problem. Therefore, since the RWA problem is NP-hard and the original RSCSA problem is more complex than the simplified one, we can conclude that the original RSCSA problem is also NP-hard.

\section{Integer linear programming model for the RSCSA problem}
As introduced in Section~III-A, the static problem is mainly related to the network planning phase and hence is not subject to strict computational time constraints. Accordingly, complex and time-consuming mathematical optimization approaches, such as ILP, can be applied to solve the static problem \cite{klinkowski2018survey}. Therefore, in this section, we propose an ILP model for the static RSCSA problem.

\subsection{Parameters}
\label{ilp-parameters}
\begin{description}
\item[$V$] the set of nodes $v$ in the network.
\item[$E$] the set of links $e$ in the network.
\item[$NP$] the set of node pairs $np$ in the network.
\item[$R$] the set of connection requests $r = \langle s_r, d_r, t_r \rangle$, where $s_r$, $d_r$, and $t_r$ represent the source node, destination node, and traffic volume [bps], respectively, of connection request $r$.
\item[${R}_{np}$] the set of connection requests between node pair $np$, which is defined as $R_{np} = \{r \in R | \langle s_r, d_r \rangle = np \}$.
\item[$P_r$] the set of $k$ candidate routing paths for connection request $r$, which is obtained using the $k$-shortest-path algorithm proposed in Ref.~\cite{christodoulopoulos2011elastic}.
\item[${P}_{np}$] the set of $k$ candidate routing paths between node pair $np$, where $P_{np} = P_{r}$ for each $r \in R_{np}$.
\item[$L$] the set of SLs $l \in L$ (per link) in the network.
\item[$L_W$] the set of SLs with wavelength switching support in the network.
\item[$L_{NW}$] the set of SLs without wavelength switching support in the network.
\item[$m^p$] the highest feasible modulation level for routing path $p$ based on its length [km].
\item[$t_{\text{\it OC}}^p$] the traffic volume [bps] that a single OC can support on routing path $p$ based on $m^p$.
\item[$F_{\text{\it GB}}$] the number of FSs occupied by an SW-GB.
\item[$F_{\text{\it OC}}$] the number of FSs occupied per OC.
\item[$F_{\text{\it max}}$] the total number of available FSs on each SL.
\item[$f_{\text{\it max}}$] the maximum index of the FSs on each SL. Note that the indices of the FSs start from 0; therefore, $f_{\text{\it max}} = F_{\text{\it max}} - 1$.
\end{description}

\subsection{Variables}
\label{ilp-variables}
\begin{description}
\item[$u^l$] a binary variable that is equal to 1 if SL $l$ is used and to 0 otherwise.
\item[$x_r^{pl}$] a binary variable that is equal to 1 if lightpath $\langle p, l \rangle$ is assigned to serve connection request $r$ and to 0 otherwise.
\item[$o_r^{pl}$] an integer variable that indicates the number of OCs that are assigned to lightpath $\langle p, l \rangle$ to serve connection request $r$.
\item[$\alpha_r^{pl}$] an integer variable that indicates the starting index of the FSs assigned to lightpath $\langle p, l \rangle$ to serve connection request $r$.
\item[$\beta_r^{pl}$] an integer variable that indicates the ending index of the FSs assigned to lightpath $\langle p, l \rangle$ to serve connection request $r$.
\item[$\theta_{rr'}^{pp'l}$] a binary variable that is equal to 1 if $\beta_r^{pl}$ is smaller than $\alpha_{r'}^{p'l}$ and to 0 otherwise.
\end{description}

\subsection{Objectives}
\begin{eqnarray}
{\mbox{Main objective:}} \quad {\mbox{\it Minimize}} && \sum_{l \in L} u^l \label{Objective1} \\
{\mbox{Minor objective:}} \quad {\mbox{\it Minimize}} && \sum_{l \in L_W} u^l \label{Objective2}
\end{eqnarray}

As stated in Section~III, the static RSCSA problem is a multiobjective problem. The main objective, shown in Eq.~(\ref{Objective1}), is to minimize the number of SLs that are used/required in the network, while the minor objective, shown in Eq.~(\ref{Objective2}), is to minimize the number of SLs with wavelength switching support that are used/required.

\subsection{Constraints}
\begin{equation}
\sum_{p \in P_r} \sum_{l \in L} t_{\text{\it OC}}^p \cdot o_r^{pl} \geq t_r \quad \forall r \in R \label{ilp-st1}
\end{equation}

For a connection request $r$, multiple lightpaths $\langle p, l \rangle$ can be established to serve it. Constraint~(\ref{ilp-st1}) ensures that the sum of the traffic volumes carried by the established lightpaths (i.e., the left-hand side) is no smaller than the required traffic volume for connection request $r$ ($t_r$).

\begin{eqnarray}
F_{\text{\it max}} \cdot x_r^{pl} \geq F_{\text{\it OC}} \cdot o_r^{pl} \quad \forall r \in R, p \in P_r, l \in L_{NW} \label{ilp-st2}
\end{eqnarray}

Constraint~(\ref{ilp-st2}) ensures that $x_r^{pl}$ is equal to 1 if there is at least one OC assigned to lightpath $\langle p, l \rangle$ to serve connection request $r$ ($o_r^{pl} \geq 1$) and is equal to 0 if no OC has been assigned ($o_r^{pl} = 0$).

\begin{equation}
\beta_r^{pl} = \alpha_r^{pl} + F_{\text{\it OC}} \cdot o_r^{pl} - x_r^{pl} \quad \forall r \in R, p \in P_r, l \in L_W  \label{ilp-st3}
\end{equation}

Constraint~(\ref{ilp-st3}) ensures the relationship between the starting and ending indices of the assigned FSs.

\begin{equation}
f_{\text{\it max}} \cdot x_r^{pl} \geq \beta_r^{pl} \quad \forall r \in R, p \in P_r, l \in L_W  \label{ilp-st4}
\end{equation}

Constraint~(\ref{ilp-st4}) ensures that if lightpath $\langle p, l \rangle$ is established to serve connection request $r$ ($x_r^{pl} = 1$), then the ending index of the FSs assigned to the lightpath ($\beta_r^{pl}$) is no greater than the maximum index of the FSs ($f_{\text{\it max}}$).

\begin{equation}
|R| \cdot k \cdot u^l \geq \sum_{r \in R} \sum_{p \in P_r} x_r^{pl} \quad \forall l \in L  \label{ilp-st5}
\end{equation}

Constraint~(\ref{ilp-st5}) ensures that $u^l$ is equal to 1 if SL $l$ has been assigned to establish at least one lightpath.

\begin{eqnarray}
&F_{\text{\it max}} \cdot (1 - x_r^{pl}) \geq F_{\text{\it OC}} \cdot o_{r'}^{p'l} \nonumber \\
&\forall r,r' \in R, p \in P_r, p' \in P_{r'}, l \in L_{NW} : p \neq p', p \bigcap p' \neq \emptyset \label{ilp-st6}
\end{eqnarray}

Constraint~(\ref{ilp-st6}) ensures that if lightpath $\langle p, l \rangle$ is established to serve connection request $r$ ($x_r^{pl} = 1$) and SL $l$ belongs to $L_{NW}$, then SL $l$ cannot be used to establish another lightpath $\langle p', l \rangle$ ($o_{r'}^{p'l} = 0$) that has one or more common links with routing path $p$ ($ p \bigcap p' \neq \emptyset$) for connection request $r'$. Note that this constraint applies only when $p \neq p'$. If this is not the case ($p = p'$), then these two lightpaths can be established on the same routing path $p$ and SL $l$ by composing an SCh of Type II (refer to the following Constraint~(\ref{ilp-st7})).

\begin{equation}
F_{\text{\it max}} \geq F_{\text{\it OC}} \cdot \sum_{r \in R_{np}} o_r^{pl} \quad np \in NP, p \in P_{np}, l \in L_{NW} \label{ilp-st7}
\end{equation}

Constraint~(\ref{ilp-st7}) indicates that for connection requests with the same source-destination pair ($r \in R_{np}$), they can be transmitted by lightpaths that share a common routing path $p \in P_{np}$ and SL $l \in L_{NW}$, composing an SCh of Type II.

\begin{eqnarray}
&\theta_{rr'}^{pp'l} + \theta_{r'r}^{p'pl} = 1 \nonumber \\
&\forall r,r' \in R, p \in P_r, p' \in P_{r'}, l \in L_W : p \bigcap p' \neq \emptyset \label{ilp-st8} \\
&\alpha_{r'}^{p'l} + F_{\text{\it max}} \cdot (1 - \theta_{rr'}^{pp'l}) \geq \beta_r^{pl} + x_r^{pl} \nonumber \\
&\forall r,r' \in R, p \in P_r, p' \in P_{r'}, l \in L_W : p = p', p \bigcap p' \neq \emptyset \label{ilp-st9} \\
&\alpha_{r'}^{p'l} + (F_{\text{\it max}} + F_{\text{\it GB}}) \cdot (1 - \theta_{rr'}^{pp'l}) \geq \beta_r^{pl} + (F_{\text{\it GB}} + 1) \cdot x_r^{pl} \nonumber \\
&\forall r,r' \in R, p \in P_r, p' \in P_{r'}, l \in L_W : p \neq p', p \bigcap p' \neq \emptyset \label{ilp-st10}
\end{eqnarray}

Constraints~(\ref{ilp-st8}) $\sim$ (\ref{ilp-st10}) ensure spectrum contiguity and spectrum nonoverlap -- the requirement of spectrum continuity is naturally satisfied -- for the lightpaths passing through SLs with wavelength switching support ($l \in L_{W}$). Since these are general constraints that have been widely applied in many previous works focusing on the static resource allocation problem in OTNs, we will not explain them in detail. However, it should be noted that two lightpaths established between the same source-destination pair can be allocated without an SW-GB in the case that their routing paths and SLs are the same (i.e., Constraint~(\ref{ilp-st9})).

\begin{equation}
F_{\text{\it max}} \cdot u^l \geq \sum_{r \in R} \sum_{p \in P_r : e \in p} F_{\text{\it OC}} \cdot o_r^{pl} \quad \forall e \in E, l \in L  \label{ilp-st11}
\end{equation}

Constraint~(\ref{ilp-st11}) is a redundant constraint. For each SL $l$ and link $e$, this constraint stipulates that the total number of FSs assigned to the lightpaths that traverse them should be no greater than $F_{\text{\it max}}$. As seen from the results of our simulation experiments, this constraint is able to significantly improve the convergence rate of the ILP model.

\section{Heuristic algorithm for solving the RSCSA problem}
\begin{figure*}[t]
	\begin{center}
		\includegraphics[width=18cm]{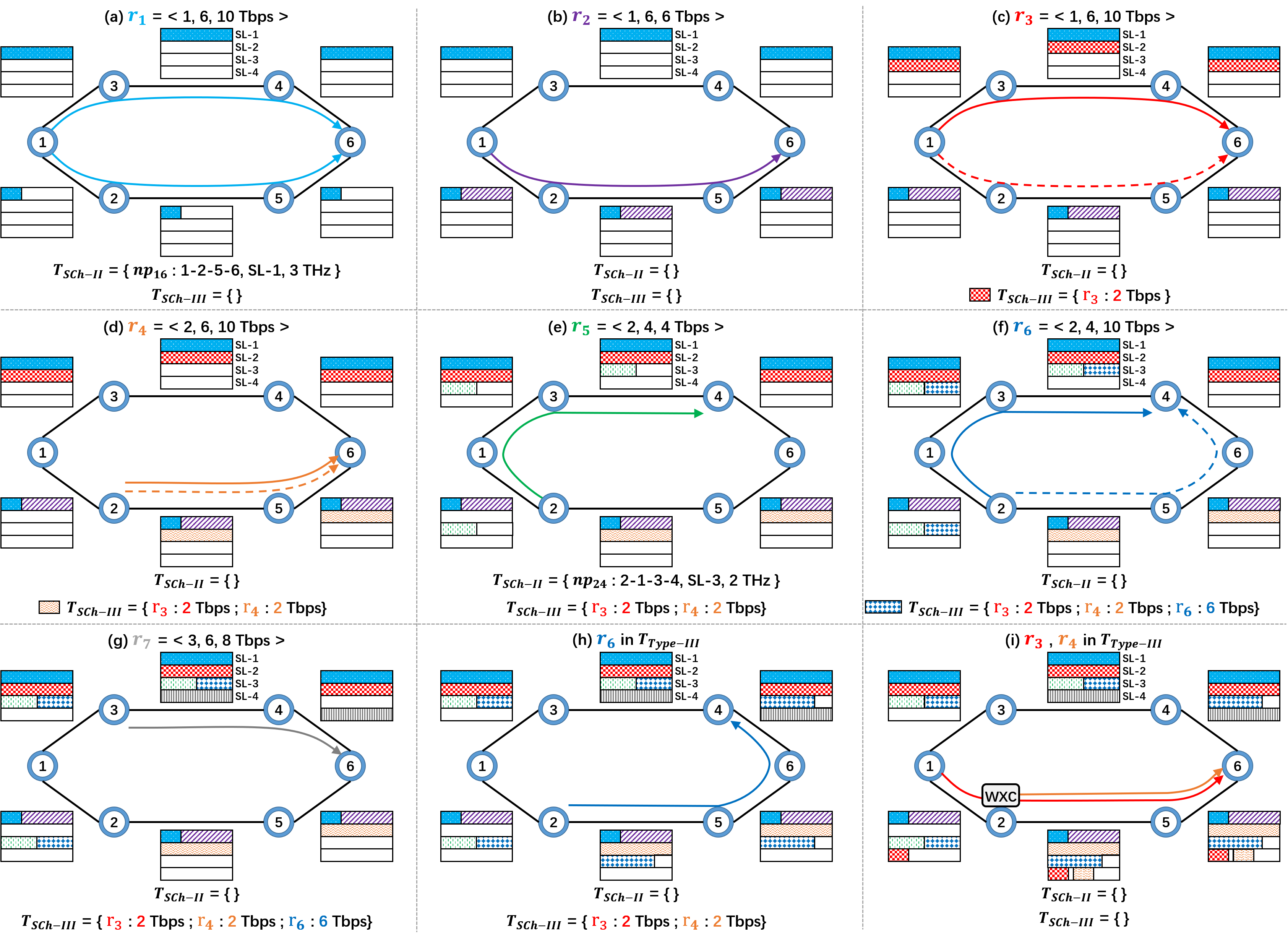}
	\end{center}
	\caption{Illustration of the proposed heuristic algorithm.}
	\label{fig-HA}
\end{figure*}
In this subsection, we propose a heuristic algorithm to solve the static RSCSA problem. First, we introduce two tables and a function used in the heuristic algorithm, as follows:
\begin{itemize}
\item We define a table $T_{\text{\it SCh-II}}$. Each entry $\langle$$np$: $p$, $l$, $B$$\rangle$ in this table records an SCh established between node pair $np$ passing through routing path $p$ and SL $l$, where the spectrum on this SCh is currently not fully used -- $B$ represents the remaining available spectrum on this SCh. In the heuristic algorithm, the connection requests are assigned resources one by one. Therefore, the remaining available spectrum on an SCh recorded in $T_{\text{\it SCh-II}}$ is expected to be assigned to subsequent connection requests with the same source-destination pair (i.e., between the same $np$) to compose an SCh of Type II.
\item We also define a table $T_{\text{\it SCh-III}}$. Each entry $\langle r: t_{\text{\it rem}}^r \rangle$ in this table records the currently unsatisfied traffic volume $t_{\text{\it rem}}^r$ for connection request $r$. The unsatisfied traffic volumes of the connection requests recorded in $T_{\text{\it SCh-III}}$ are expected to be served by SChs of Type III.
\item We define a function named {\it First-Fit SL Allocation (FF-SLA)}. This function takes a routing path $p$ as input and determines the available SL with the lowest index along $p$ (denoted by $l_{\text{\it FF}}^p$). In this paper, we assume that the indices of SLs without wavelength switching support (i.e., $l \in L_{NW}$) are lower than those of SLs with wavelength switching support (i.e., $l \in L_{W}$). The output of this function is $\langle$$p$, $l_{\text{\it FF}}^p$$\rangle$.
\end{itemize}

The heuristic algorithm is divided into three parts, and we will explain each of them individually. To facilitate readers' understanding, we present a simple illustration in Fig.~\ref{fig-HA}. We consider a 6-node network with 4 SLs (per link), of which only one, SL-4, supports wavelength switching. For simplicity, we assume that each SL has 4 THz of available spectrum and supports 8 Tbps of traffic volume regardless of the path length (i.e., without considering adaptive modulation). Then, we consider a set of connection requests $R$ = \{$r_1$, $r_2$, $\cdots$, $r_7$\}. These connection requests belong to different sets $R_{np}$: $R_{np_{16}}$ = \{$r_1$, $r_2$, $r_3$\}, $R_{np_{26}}$ = \{$r_4$\}, $R_{np_{24}}$ = \{$r_5$, $r_6$\}, and $R_{np_{36}}$ = \{$r_7$\}. The connection requests will be assigned resources one by one following a specified service sequence $R_{\text{\it seq}}$.

\subsection{Assignment for SChs of Type I and Type II}
Initially, we attempt to assign SChs of Type I and Type II for all connection requests. Fig.~\ref{fig-HA}.(a) shows the arrival of the first connection request, $r_1 = \langle 1, 6, 10 \: \text{Tbps} \rangle$. According to the output of the {\it FF-SLA} function, SL-1 and routing path $\langle$1-3-4-6$\rangle$, with the shortest distance, are first selected to establish an SCh of Type I with support for 8 Tbps of traffic volume. However, 2 Tbps of the traffic volume of the request still needs to be satisfied. Therefore, by running the {\it FF-SLA} function again, SL-1 and routing path $\langle$1-2-5-6$\rangle$ are additionally selected, and another SCh is established. Since the spectrum available on this SCh is not fully used, the entry $\langle$$np_{16}$: 1-2-5-6, SL-1, 3 THz$\rangle$ is appended to $T_{\text{\it SCh-II}}$. The remaining 3 Thz of spectrum is expected to be assigned to the subsequent connection requests $r_2$ and $r_3$, which belong to the same $R_{np_{16}}$ as $r_1$, to compose an SCh of Type II. Then, we remove $r_1$ from $R_{np_{16}}$.

Subsequently, a connection request $r_2 = \langle 1, 6, 6 \: \text{Tbps} \rangle$ arrives, as shown in Fig.~\ref{fig-HA}.(b). The remaining available spectrum on the SCh recorded in $T_{\text{\it SCh-II}}$ has the highest priority for assignment to subsequent connection requests. Therefore, we first check whether there is an SCh between $np_{16}$ that is not fully used recorded in $T_{\text{\it SCh-II}}$. In this case, it is obvious that $r_2$ can be transmitted using the remaining 3 THz of spectrum on the SCh recorded in $T_{\text{\it SCh-II}}$ above by composing an SCh of Type II. Since this SCh is fully used after being assigned to $r_2$, the corresponding entry is removed from $T_{\text{\it SCh-II}}$. Finally, we remove $r_2$ from $R_{np_{16}}$.

Then, a connection request $r_3 = \langle 1, 6, 10 \: \text{Tbps} \rangle$ arrives, as shown in Fig.~\ref{fig-HA}.(c). We call the {\it FF-SLA} function because there is no SCh between $np_{16}$ recorded in $T_{\text{\it SCh-II}}$ at this time. Thus, SL-2 and routing path $\langle$1-3-4-6$\rangle$ are selected to establish an SCh of Type I. Similar to the case of $r_1$, 2 Tbps of the traffic volume of the request remains to be satisfied. However, we will not establish another not fully used SCh in this case because if we were to establish such an SCh (represented by the red dotted line), its remaining available spectrum would have no chance to be used because no subsequent connection request exists between $np_{16}$, and this would result in a waste of spectrum on this SCh. Instead, an entry $\langle$$r_3$: 2 Tbps$\rangle$ is appended to $T_{\text{\it SCh-III}}$. The unsatisfied 2 Tbps of traffic volume for $r_3$ is expected to be served by an SCh of Type III -- sharing the spectrum with other connection requests between different source-destination pairs.

The procedures described above will be repeated for each connection request $r \in R$ (e.g., Fig.~\ref{fig-HA}.(a)$\sim$(g) for this example). Notably, SChs of Type I and Type II will not result in any spectrum fragmentation and offer SW-GB savings compared to SChs of Type III. Therefore, SChs of Type I and Type II always have a higher priority for establishment than SChs of Type III. Consequently, although it is preferable to use SLs without wavelength switching support (i.e., $l \in L_{NW}$) when establishing SChs of Type I and Type II (this is the reason why lower indices are assigned to the SLs without wavelength switching), SLs with wavelength switching support (i.e., $l \in L_{W}$) are also allowed to be used if the available SLs without wavelength switching support are inadequate (see, for example, the assignment of $r_7$ in Fig.~\ref{fig-HA}.(g)). The pseudocode for this part of the algorithm is shown in Algorithm~\ref{Code-HA1}.
\begin{algorithm}[!htbp]
	\footnotesize
	\caption{Assignment for SChs of Type I and Type II}
	\label{Code-HA1}
	\begin{algorithmic}[1]
		\REQUIRE $R_{\text{\it seq}}$, $R_{np}$ for each $np \in NP$
		\ENSURE $T_{\text{\it SCh-III}}$
		\STATE Create new tables: $T_{\text{\it SCh-II}}$ and $T_{\text{\it SCh-III}}$.
		\FOR {each $r = \langle s, d, t \rangle \in R_{\text{\it seq}}$}
		\STATE Remove $r$ from $R_{np_{\text{\it sd}}}$.
		\IF {an SCh for $\langle$$np_{\text{\it sd}}$: $p$, $l$, $B$$\rangle$ is recorded in $T_{\text{\it SCh-II}}$}
		\STATE $t_p$ $\leftarrow$ calculate the supportable traffic volume on the SCh based on the highest feasible modulation format $m_p$ for routing path $p$ and the remaining available spectrum $B$.
		\IF {$t_p > t$}
		\STATE Assign routing path $p$, SL $l$, and the required spectrum to $r$ -- create a (not fully used) SCh of Type II.
		\STATE $t$ $\leftarrow 0$
		\STATE $B$ $\leftarrow$ $B$ minus the required spectrum for $r$.
		\STATE Go to the next connection request (line 2).
		\ELSE
		\STATE Assign routing path $p$, SL $l$, and the remaining available spectrum $B$ to $r$ -- create an SCh of Type II.
		\STATE $t$ $\leftarrow$ $t - t_p$.
		\STATE Remove $\langle$$np_{\text{\it sd}}$: $p$, $l$, $B$$\rangle$ from $T_{\text{\it SCh-II}}$.
		\ENDIF
		\ENDIF
		\WHILE {$t > 0$}
		\STATE $\langle${\it best}-$p_r$, {\it best}-$l_r$$\rangle$ $\leftarrow$ call the {\it FF-SLA} function for each candidate path $p_r \in P_r$ and select the one with the smallest $l_{\text{\it FF}}^{p_r}$.
		\STATE $t_{\text{\it best-}p_r}$  $\leftarrow$ calculate the supportable traffic volume on routing path {\it best}-$p_r$ and SL {\it best}-$l_r$ based on the highest feasible modulation format for {\it best}-$p_r$.
		\IF {$t_{\text{\it best-}p_r} > t$}
		\IF {$R_{np_{\text{\it sd}}}$ is not an empty set}
		\STATE Assign routing path {\it best}-$p_r$, SL {\it best}-$l_r$, and the required spectrum to $r$ -- create a (not fully used) SCh.
		\STATE $B_{\text{\it rem}}$ $\leftarrow$ calculate the remaining available spectrum of the SCh.
		\STATE Append $\langle$$np_{\text{\it sd}}$: {\it best}-$p_r$, {\it best}-$l_r, B_{\text{\it rem}}$$\rangle$ to $T_{\text{\it SCh-II}}$.
		\ELSE
		\STATE Append $\langle$$r$: $t$$\rangle$ to $T_{\text{\it SCh-III}}$.
		\ENDIF
		\STATE $t$ $\leftarrow 0$
		\ELSE
		\STATE Assign routing path {\it best}-$p_r$, SL {\it best}-$l_r$, and the entire C-band spectrum to $r$ -- create an SCh of Type I.
		\STATE $t$ $\leftarrow$ $t - t_{\text{\it best-}p_r}$.
		\ENDIF
		\ENDWHILE
		\ENDFOR
		\RETURN $T_{\text{\it SCh-III}}$
	\end{algorithmic}
\end{algorithm}

\subsection{Reassignment for SChs of Type I and Type II}
Using Algorithm~\ref{Code-HA1}, we have assigned SChs of Type I and Type II to each connection request and obtained a table containing a set of connection requests with currently unsatisfied traffic volumes (i.e., $T_{\text{\it SCh-III}}$), which are expected to be served by SChs of Type III. Here, $l_{\text{\it NW-A1}}^{\text{\it max}}$ denotes the maximum index of the currently used/required SLs without wavelength switching support (i.e., the number of such SLs) in the network after the execution of Algorithm~\ref{Code-HA1}. It is obvious that $l_{\text{\it NW-A1}}^{\text{\it max}}$ SLs without wavelength switching support may not be used on every link. For example, as shown in Fig.~\ref{fig-HA}.(g), $l_{\text{\it NW-A1}}^{\text{\it max}}$ is equal to 3, but only 2 SLs are used on link 2-5. Note that these unused SLs (e.g., SL-3 on link 2-5) cannot be used to establish SChs of Type III hereafter because they do not support wavelength switching. Therefore, before we assign SChs of Type III to the connection requests recorded in $T_{\text{\it SCh-III}}$, we will first attempt to assign them one by one -- starting from the one with the largest unsatisfied traffic volume -- to the unused SLs whose indices are smaller than $l_{\text{\it NW-A1}}^{\text{\it max}}$. As shown in Fig.~\ref{fig-HA}.(h), the connection request $r_6$ can be successfully assigned to pass through routing path $\langle$2-5-6-4$\rangle$, although this will result in a certain degree of spectrum wastage. Then, we remove $\langle$$r_6$: 6 Tbps$\rangle$ from $T_{\text{\it SCh-III}}$. In this way, we can somewhat reduce the number of entries in $T_{\text{\it SCh-III}}$, thus making it possible to use fewer SLs with wavelength switching support hereafter. Such an assignment will not result in any negative effect on the optimization objective(s) because the (main) objective of the RSCSA problem is to minimize the number of SLs that are used/required in the network, not to minimize their sum over all links.

The pseudocode for this part of the algorithm is shown in Algorithm~\ref{Code-HA2}. The inputs to Algorithm~\ref{Code-HA2} are $T_{\text{\it SCh-III}}$ and $l_{\text{\it NW-A1}}^{\text{\it max}}$, which are obtained after the execution of Algorithm~\ref{Code-HA1}. The output of Algorithm~\ref{Code-HA2} is the modified $T_{\text{\it SCh-III}}$, in which the number of entries may be reduced.
\begin{algorithm}[!htbp]
	\footnotesize
	\caption{Reassignment for SChs of Type I and Type II}
	\label{Code-HA2}
	\begin{algorithmic}[1]
		\REQUIRE $T_{\text{\it SCh-III}}$, $l_{\text{\it NW-A1}}^{\text{\it max}}$
		\ENSURE $T_{\text{\it SCh-III}}$
		\STATE Sort $T_{\text{\it SCh-III}}$ by $t_{\text{\it rem}}^r$, from largest to smallest.
		\FOR {each $\langle r: t_{\text{\it rem}}^r \rangle$ in $T_{\text{\it SCh-III}}$}
		\WHILE {TRUE}
		\STATE $\langle${\it best}-$p_r$, {\it best}-$l_r$$\rangle$ $\leftarrow$ call the {\it FF-SLA} function for each candidate path $p_r \in P_r$, and select the one with the smallest $l_{\text{\it FF}}^{p_r}$.
		\IF {{\it best}-$l_r \leq l_{NW-A1}^{\text{\it max}}$}
		\STATE $t_{\text{\it best-}p_r}$  $\leftarrow$ calculate the supportable traffic volume on routing path {\it best}-$p_r$ and SL {\it best}-$l_r$ based on the highest feasible modulation format for {\it best}-$p_r$.
		\IF {$t_{\text{\it best-}p_r} > t_{\text{\it rem}}^r$}
		\STATE Assign routing path {\it best}-$p_r$, SL {\it best}-$l_r$, and the required spectrum to $r$ -- create a (not fully used) SCh of Type I.
		\STATE Remove $\langle r: t_{\text{\it rem}}^r \rangle$ from $T_{\text{\it SCh-III}}$.
		\STATE \textbf{break while} - go to the next connection request (line 2).
		\ELSE
		\STATE Assign routing path {\it best}-$p_r$, SL {\it best}-$l_r$, and the entire C-band spectrum to $r$ -- create an SCh of Type I.
		\STATE $t_{\text{\it rem}}^r$ $\leftarrow$ $t_{\text{\it rem}}^r - t_{\text{\it best-}p_r}$.
		\ENDIF
		\ELSE
		\STATE \textbf{break while} - go to the next connection request (line 2).
		\ENDIF
		\ENDWHILE
		\ENDFOR
	\end{algorithmic}
\end{algorithm}

\subsection{Assignment for SChs of Type III}
Finally, we begin to assign resources to the unsatisfied connection requests that are still recorded in $T_{\text{\it SCh-III}}$ after Algorithm~\ref{Code-HA2} has been executed. Similar to the approach that has been widely applied to the previous RSA and RSSA problems, each connection request will be assigned using the {\it First-Fit Spectrum Allocation (FF-SA)} function \cite{christodoulopoulos2011elastic}, as shown in Fig.~\ref{fig-HA}.(i). The pseudocode for this part of the algorithm is shown in Algorithm~\ref{Code-HA3}, where $l_{W}^{\text{\it min}}$ and $l_{W}^{\text{\it max}}$ represent the minimum and maximum indices, respectively, of SLs with wavelength switching support (i.e., $l \in L_W$). As stated before, the indices of the SLs with wavelength switching support (i.e., $l \in L_{NW}$) are lower than those of the SLs without wavelength switching support. Therefore, $l_{W}^{\text{\it min}}$ and $l_{W}^{\text{\it max}}$ are actually equal to $|L_{NW}| + 1$ and $|L|$, respectively.
\begin{algorithm}[!htbp]
	\footnotesize
	\caption{Assignment for SChs of Type III}
	\label{Code-HA3}
	\begin{algorithmic}[1]
		\REQUIRE $T_{\text{\it SCh-III}}$
		\ENSURE $T_{\text{\it SCh-III}}$
		\STATE $l_W^{\text{\it current}}$ $\leftarrow$ $l_W^{\text{\it min}}$
		\STATE Sort $T_{\text{\it SCh-III}}$ by $t_{\text{\it rem}}^r$, from largest to smallest.
		\WHILE {$l_W^{\text{\it current}} \leq l_W^{\text{\it max}}$ \textbf{and} $T_{\text{\it SCh-III}}$ is not empty}
		\FOR {each $\langle r: t_{\text{\it rem}}^r \rangle$ in $T_{\text{\it SCh-III}}$}
		\STATE {\it best}-$p_r$ $\leftarrow$ apply the {\it FF-SA} function \cite{christodoulopoulos2011elastic} for each candidate path $p_r \in P_r$ on SL $l_W^{\text{\it current}}$ and select the one with the lowest ending index of FSs.
		\IF {{\it best}-$p_r$ $\neq$ None}
		\STATE Assign routing path {\it best}-$p_r$, $l_W^{\text{\it current}}$, and the required FSs as obtained by the {\it FF-SA} function to $r$.
		\STATE Remove $\langle r: t_{\text{\it rem}}^r \rangle$ from $T_{\text{\it SCh-III}}$.
		\ENDIF
		\ENDFOR
		\STATE $l_W^{\text{\it current}}$ $\leftarrow$ $l_W^{\text{\it current}}$ + 1
		\ENDWHILE
		\IF{$T_{\text{\it SCh-III}}$ is not empty}
		\STATE Call Algorithm~\ref{Code-HA2} again while allowing {\it best}-$l_r > l_{\text{\it NW-A1}}^{\text{\it max}}$ in line 5.
		\ENDIF
	\end{algorithmic}
\end{algorithm}

Notably, we may not be able to successfully serve all unsatisfied connection requests recorded in $T_{\text{\it SCh-III}}$ if the available SLs with wavelength switching support are inadequate. In this case, we will call Algorithm~\ref{Code-HA2} again, now allowing the use of SLs with indices greater than $l_{\text{\it NW-A1}}^{\text{\it max}}$ (i.e., removing line 5 and the corresponding lines 15 $\sim$ 17 from Algorithm~\ref{Code-HA2}).

\subsection{Iteration with the simulated annealing metaheuristic}
Similar to previous works focusing on the static RWA, RSA, and RSSA problems, the service sequence $R_{\text{\it seq}}$ is very important to our heuristic algorithm for solving the RSCSA problem because the heuristic algorithm assigns resources to the connection requests one by one. Different service sequences will lead to different assignment results. Therefore, we apply the simulated annealing (SimAn) metaheuristic approach \cite{christodoulopoulos2011elastic} to find a good sequence that yields better results. Undoubtedly, other iterative approaches, such as simple random shifting, could also be applied for this purpose.

\section{Simulations and performance evaluations}
\begin{figure}[!htbp]
	\begin{center}
		\includegraphics[width=7cm]{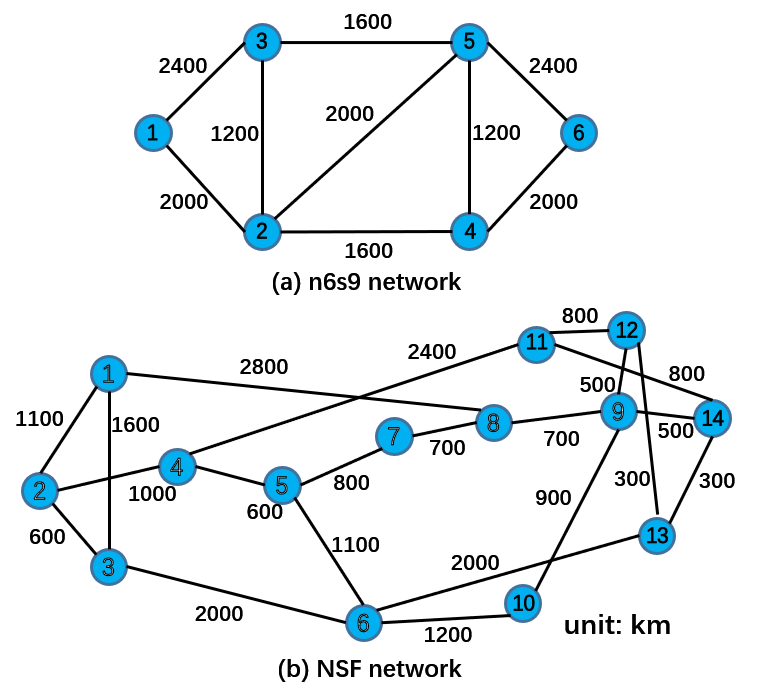}
	\end{center}
	\caption{Network topologies: (a) the simple 6-node, 18-directed-link n6s9 network; (b) the realistic 14-node, 42-directed-link NSF network \cite{yang2018routing}.}
	\label{fig-nets}
\end{figure}
In this section, we evaluate the performance of the proposed ILP model and heuristic algorithm based on two network topologies: i) the simple 6-node, 18-directed-link n6s9 network, as shown in Fig.~\ref{fig-nets}.(a), and ii) the realistic 14-node, 42-directed-link NSF network, as shown in Fig.~\ref{fig-nets}.(b) \cite{yang2018routing}. The following assumptions are adopted in the simulation experiments:

\begin{itemize}
\item A bundle of weakly coupled 4-core multicore fibers (MCFs), as proposed in Ref.~\cite{matsui2015design}, is assumed for each link (i.e., $|L| = 40$) in the networks for the following reasons: i) full compatibility with conventional SMFs while maintaining a 125 {\it $\mu$m} cladding diameter; ii) low intercore crosstalk (XT), enabling ultralong-haul all-optical transmission; and iii) significant cost savings when combined with the application of cladding-pumped multicore erbium-doped fiber amplifiers (MC-EDFAs) \cite{abedin2012cladding, rivas2016cost}.

\item The total spectrum per core of a 4-core MCF is considered to be 4 THz (C-band), that is, 320 FSs conforming to the ITU-T 12.5 GHz grid \cite{ITU-T}.

\item Each subtransceiver operates at a fixed baud rate of 32 Gbaud, supporting an OC that occupies 37.5 GHz (i.e., 3 FSs) \cite{jinno2019spatial, khodashenas2016comparison}. In addition, a spectrum occupation of 12.5 GHz (i.e., 1 FS) is assumed for each SW-GB.

\item We consider four modulation formats in the simulation experiments, namely, double polarization (DP) BPSK, QPSK, 8-QAM, and 16-QAM. The supportable bit rates per OC are 50, 100, 150 and 200 Gbps. Notably, the transmission reaches for the different modulation formats are bounded by two factors: i) the optical signal-to-noise ratio (OSNR) and ii) the XT \cite{perell2016flex}. However, since we consider 4-core MCFs with low XT interference, the transmission reaches are mainly bounded by the OSNR in this case. Therefore, for the different modulation formats listed above, the transmission reaches are 6300, 3500, 1200, and 600 km, respectively \cite{khodashenas2016comparison}.

\item Three candidate shortest routing paths ($k$ = 3) are considered for each connection request.
\end{itemize}

Moreover, to compare the network performance of SDM-OTNs and SCNs, we consider three different OXCs, as follows:

\begin{itemize}
\item The first is the conventional OXC applied in SDM-OTNs, which is implemented using stacked WXCs as the basic solution to achieve SDM. In such an OXC, wavelength switching is supported on each SL -- i.e., $L_W = L$ and $L_{NW} = \emptyset$.

\item The second is an HOXC (i.e., SXC+WXC) proposed for SCNs, which is implemented using CSSs as shown in Fig.~\ref{fig-SCN}. We assume that in such an HOXC, one-ninth of the SLs support wavelength switching, in accordance with the assumptions proposed in Ref.~\cite{jinno2019spatial}. That is, $|L_W| = \lceil {|L| \over 9} \rceil$ and $L_{NW} = L - L_W$.

\item The last is an OXC that does not support wavelength switching on any of its SLs. In such an OXC, only SXCs are deployed at intermediate nodes -- i.e., $L_W = \emptyset$ and $L_{NW} = L$.
\end{itemize}

All simulation experiments were performed in a Microsoft Windows 10 environment using a computer with an AMD Ryzen 6-core 3.6 GHz CPU and 16 GB of memory.

\subsection{Simulation experiments involving the simple n6s9 network}
\begin{figure*}[!htbp]
\begin{center}
		\includegraphics[width=18cm]{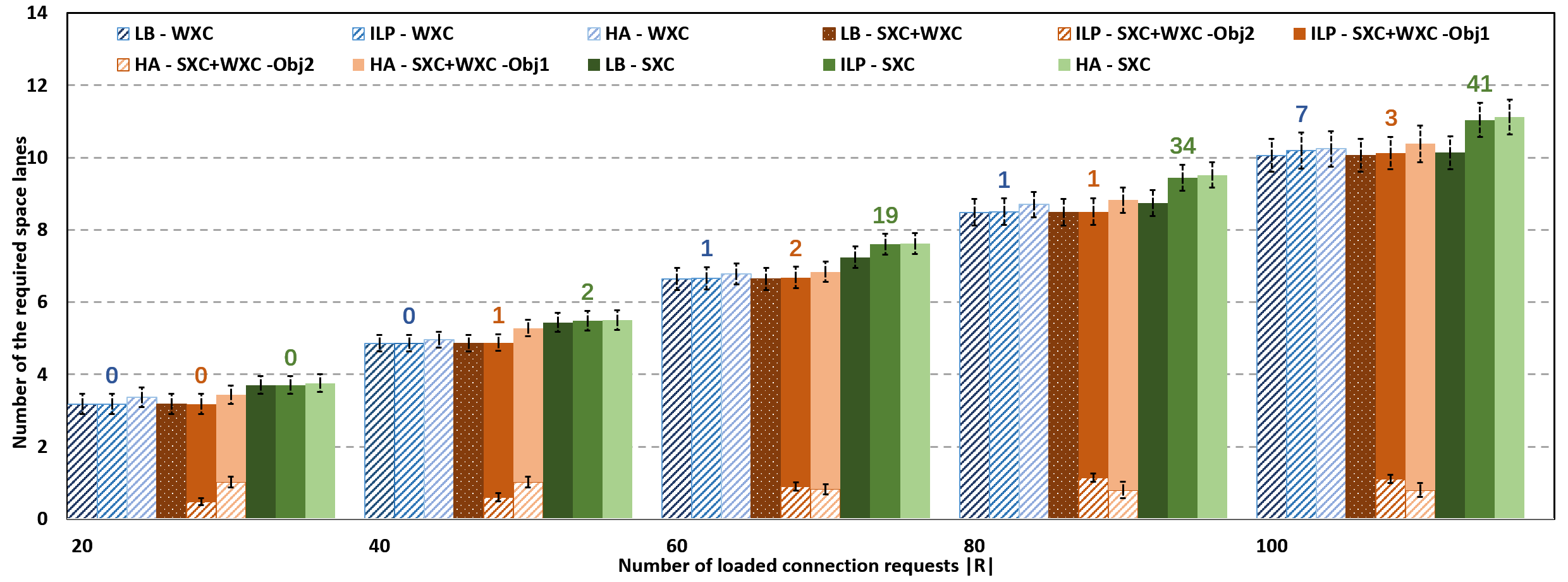}
\end{center}
\caption{Simulation results for the simple 6-node, 18-link n6s9 network.}
	\label{fig-data-n6}
\end{figure*}
In these simulation experiments, we considered the simple, small-scale n6s9 network with 20 SLs (i.e., one bundle of five 4-core MCFs per link). Therefore, in the HOXC case, the set of SLs with wavelength switching support, $L_W$, was $\{ 18, 19, 20 \}$. We considered different numbers of connection requests ranging from 20 to 100 (in increments of 20), representing different traffic loads. Specifically, the total average traffic volumes with which the network was loaded ranged from 0.11 to 0.55 Pbps. For each traffic load, we randomly generated 50 different traffic matrices $R$. Considering that current traffic volumes are expected to increase by 10$\times$ in the future (by 2024) \cite{winzer2017scaling, jinno2019spatial}, for each unidirectional connection request in $R$, the traffic volume was randomly selected from among traffic profiles of \{1 Tbps, 4 Tbps, 10 Tbps\} with probabilities of \{0.3, 0.3, 0.4\} \cite{rumipamba2018on, rumipamba2018space, talebi2017on, neto2018ccaling, moura2018routing}.

To solve the ILP model proposed in Section~IV, we used the optimization software {\it GUROBI v8.0.1} \cite{gurobi}. Since the RSCSA problem is an NP-hard problem, as proven in Section III-B, it may not be possible to completely solve the ILP model within a reasonable amount of time for certain input matrices and/or traffic loads. Therefore, we bounded the running time of the ILP model to 1 hour for the main objective and 300 seconds for the minor objective. Moreover, the solutions given by the heuristic algorithm were input into the ILP model as initial solutions to improve the convergence rate.

The simulation results, including the average values of the objective(s) and the 95\% confidence intervals ({\it T-distribution}), are shown in Fig.~\ref{fig-data-n6}. The abbreviations `LB', `ILP', and `HA' in Fig.~\ref{fig-data-n6} represent the lower bound of the RSCSA problem given by the `{\it BestBound}' of {\it GUROBI}, the optimal or current feasible solution obtained by solving the ILP model with a 1-hour running time limit, and the solution obtained using the heuristic algorithm with 1000 iterations of $R_{\text{\it seq}}$, respectively. The abbreviations `WXC', `SXC+WXC', and `SXC' represent the three OXCs introduced above, that is, the OXC with full wavelength switching support (i.e., $L_{W} = L$) for SDM-OTNs, the HOXC with partial wavelength switching support (i.e., $|L_W| = 3$) for SCNs, and the OXC without wavelength switching support (i.e., $L_{W} = \emptyset$), respectively. Moreover, the number over the data bar represents the number of input matrices $R$ for which the corresponding ILP models did not yield optimal solutions within 1 hour.

From Fig.~\ref{fig-data-n6}, we can see that even though only approximately one-ninth of the SLs support wavelength switching in the HOXC case, the results of `ILP - SXC+WXC' and `ILP - WXC' are the same, while negligible gaps (within 2.4\%) exist between the results of `HA - SXC+WXC' and `HA - WXC'. Moreover, as we can see from the results of `ILP - SXC+WXC - Obj2' and `HA - SXC+WXC - Obj2', the average numbers of used/required SLs with wavelength switching support for the solutions obtained using both the ILP model and the heuristic algorithm are less than 1.2 for all traffic loads in the HOXC case. These observations indicate that the conventional OXC with full wavelength switching support offers no remarkable advantages for future connection requests with large traffic volumes (e.g., several or dozens of Tbps) -- or, equivalently, for multiple connection requests between the same source-destination pair with smaller traffic volumes typical of current network traffic that are groomed into a single connection request with a larger traffic volume. Moreover, according to the cost assessments presented in Refs.~\cite{asano2018cost} and \cite{jinno2019spatial}, for the network with 20 SLs considered in these simulation experiments, the cost of either a full-size CSS-based HOXC or a sub-CSS-based HOXC (see Fig.~\ref{fig-SCN}.(a)) designed for SCNs is only 25\% of that of a conventional OXC with full wavelength switching support designed for SDM-OTNs. Therefore, full wavelength switching support may no longer be necessary for the future massive SDM era.

In contrast, relatively large gaps, ranging from 8\% to 14\%, can be observed between the results for OXCs without wavelength switching support (i.e., `SXC') and those for the above two (H)OXC cases with full/partial wavelength switching support. These findings indicate that completely removing wavelength switching support from the intermediate nodes will result in some loss of network performance. However, fewer cost savings (compared with the great cost savings between `WXC' and `SXC+WXC') can be achieved, as well. The trade-off decision should be made by the network operators.

Moreover, we can observe that the `ILP' and `HA' results are very similar in all cases. For the two (H)OXC cases with full/partial wavelength switching support (i.e., `WXC' and `SXC+WXC'), the ILP model can be completely solved within 1 hour for all or the majority of the input matrices $R$, depending on the traffic loads, and the results of both `ILP' and `HA' are close to the lower bounds of the problem. For the OXC case without wavelength switching support (i.e., `SXC'), the ILP model becomes difficult to solve within 1 hour if the traffic load is heavy. In this case, the gaps between the `HA' results and the lower bounds range from 1.1\% to 8.8\%, while those between the `ILP' results and the lower bounds range from 0.7\% to 8.2\%, which are considered acceptable.
\begin{table}[!htbp]
	\scriptsize
	\centering
	\caption{Average running times of the proposed heuristic algorithm with 1000 iterations for the simple n6s9 network}
	\setlength{\tabcolsep}{7pt}
	\label{tab-time-n6}
	\begin{tabular}{c | rrrrr}
		\toprule
		OXC & \multicolumn{5}{c}{Traffic load $|R|$} \\ 
		\cline{2-6} 
		\specialrule{0em}{0pt}{1pt} 
		Architecture &  20  &  40  &  60  & 80  & 100  \\ 
		\midrule
		WXC & 11.95	& 14.03	& 15.81	& 17.06	& 18.49 \\
		SXC+WXC & 3.98	& 4.09	& 4.44	& 4.20	& 4.64 \\
		SXC & 0.67	& 1.05	& 1.43	& 1.79	& 2.17 \\
		\bottomrule
	\end{tabular}
\end{table}

Table~\ref{tab-time-n6} lists the average running times (in seconds) of the heuristic algorithm with 1000 iterations (on a single thread) for the simple n6s9 network. We can see that the running times of the heuristic algorithm in the conventional OXC case with full wavelength switching are much longer than those in the HOXC case with partial wavelength switching, and the shortest running times are incurred in the OXC case without wavelength switching. The reason for this observation is that finding a set of continuous and contiguous FSs with the lowest ending index along a routing path by means of the {\it FF-SA} function is much more difficult than finding a feasible SL with the lowest index along a routing path by means of the {\it FF-SLA} function. Therefore, in the conventional OXC case with full wavelength switching, the {\it FF-SA} function will be called more times -- for each SL with wavelength switching support until all connection requests have been served -- by the heuristic algorithm, resulting in a longer running time. In contrast, in the OXC case without wavelength switching, the heuristic algorithm will not call the {\it FF-SA} function even once, since there are no SLs that support wavelength switching, resulting in the shortest running time.

In summary, the simulation results show that the proposed ILP model (with a 1-hour running time limit) and heuristic algorithm both work well for small-scale problem instances, for which the optimal solutions or solutions close to the lower bounds can be obtained.

\subsection{Simulation experiments involving the realistic NSF network}
\begin{figure}[!htbp]
\begin{center}
		\includegraphics[width=8.5cm]{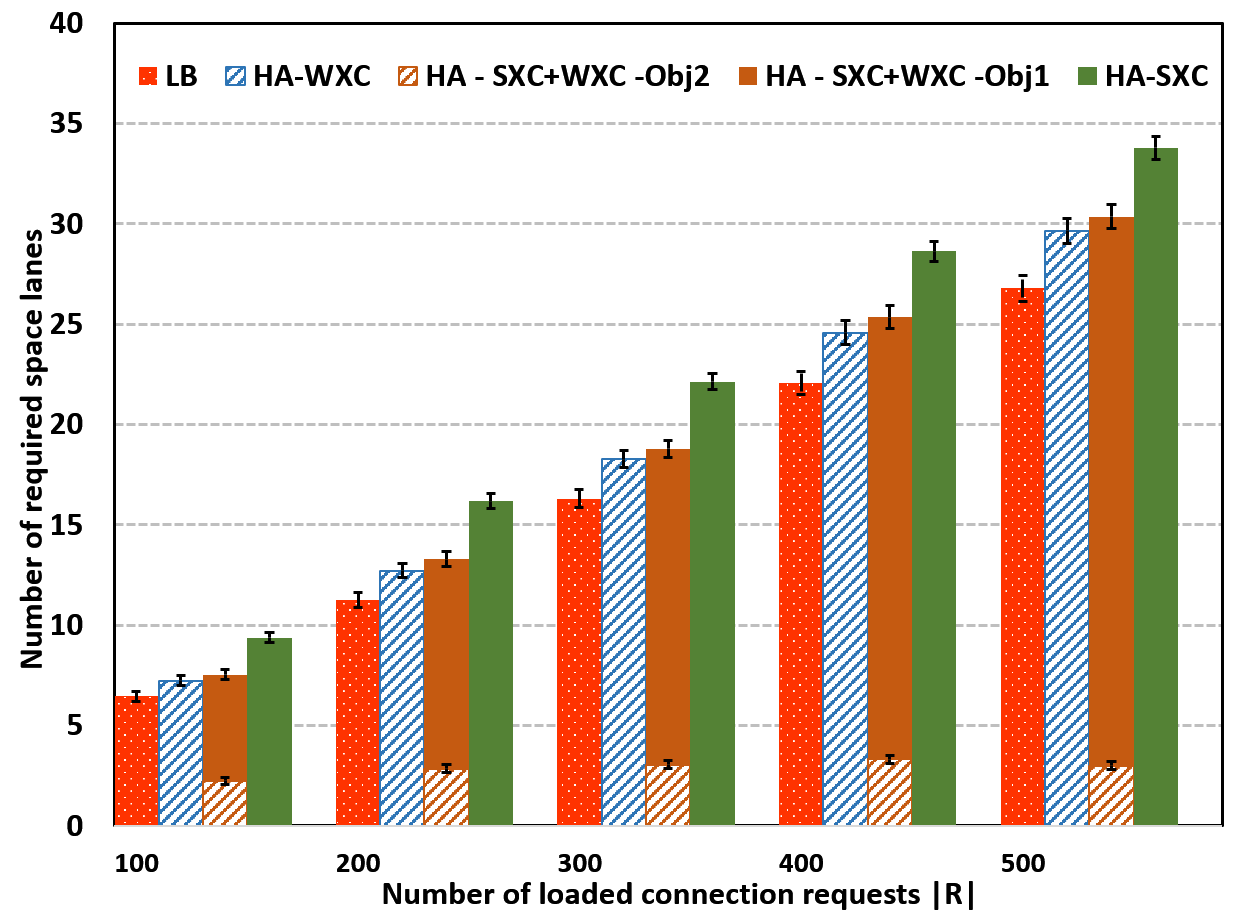}
\end{center}
\caption{Simulation results for the 14-node, 42-link NSF network.}
	\label{fig-data-nsf}
\end{figure}
In these simulation experiments, we considered the realistic large-scale NSF network with 40 SLs (i.e., one bundle of five 4-core MCFs per link). Considering that one-ninth of the SLs support wavelength switching \cite{jinno2019spatial}, the set $L_W$ was $\{ 36, 37, \cdots, 40 \}$ in this case. Moreover, we also considered heavier traffic loads -- ranging from 100 to 500 (in increments of 100) connection requests -- and 50 different traffic matrices $R$ for each traffic load. In this case, the total average traffic volumes with which the network was loaded ranged from 0.55 to 2.75 Pbps. In such large-scale instances, acceptable solutions become difficult to obtain within a reasonable amount of time by solving the ILP model. Therefore, we relaxed the original ILP model by removing Constraints~(\ref{ilp-st2}) $\sim$ (\ref{ilp-st10}) and the minor objective to obtain the lower bounds for the RSCSA problem, which we then used as the benchmarks to evaluate the performance of the heuristic algorithm. This relaxation means that i) wavelength switching is allowed on all SLs, ii) lightpaths can be established without SW-GBs, and iii) the spectrum contiguity constraint is relaxed. Consequently, in this case, the lower bound obtained by solving the relaxed ILP model is not only the lower bound of the RSCSA problem in an SCN but also the lower bound of the RSSA problem in an SDM-OTN -- if we transform the objective of the RSSA problem into the minimization of the number of SLs, as opposed to the number of FSs, that are used/required in the network.

The corresponding simulation results are shown in Fig.~\ref{fig-data-nsf}. We can observe that the results in Fig.~\ref{fig-data-nsf} are similar to those presented in Fig.~\ref{fig-data-n6}. First, the gaps between the results of `HA - SXC+WXC' and `HA - WXC' are negligible, ranging from 2.3\% to 4.2\%. This means that the conventional OXC with full wavelength switching support is not a preferred solution for future Pbps-level OTNs because of the significantly higher cost -- for the network with 40 SLs considered here, the conventional OXC configuration is 5.8 times as costly as the full-size CSS-based HOXC configuration and 4.2 times as costly as the sub-CSS-based HOXC configuration \cite{asano2018cost, jinno2019spatial} -- for similar performance. By contrast, we can see that the gaps between the results of `HA - SXC+WXC' and `HA - SXC' are relatively significant, ranging from 10.1\% to 19.6\% for different traffic loads. Therefore, the network operators are required to make a decision concerning the balance between the additional cost and better performance.

Moreover, the results of `HA - WXC' are close to the lower bounds obtained by solving the relaxed ILP model (i.e., `LB'). The gaps between them range from 9.6\% to 11.4\%. Compared to the results shown in Fig.~\ref{fig-data-n6}, these gaps are relatively large because the lower bounds for these simulation experiments are not strict -- they are obtained by solving the relaxed ILP model, in which almost all of the constraints of the original ILP model have been removed. In addition, it should be noted that it is unfair to evaluate the performance of the heuristic algorithm by comparing the results of `HA - SXC+WXC' or `HA - SXC' against these lower bounds because wavelength switching is allowed on all SLs in the relaxed ILP model.
\begin{table}[t]
	\scriptsize
	\centering
	\caption{Average running times of the proposed heuristic algorithm with 1000 iterations for the realistic NSF network}
	\setlength{\tabcolsep}{6pt}
	\label{tab-time-nsf}
	\begin{tabular}{c | rrrrr}
		\toprule
		OXC & \multicolumn{5}{c}{Traffic load $|R|$} \\
		\cline{2-6} 
		\specialrule{0em}{0pt}{1pt} 
		Architecture &  100  &  200  &  300  & 400  & 500  \\ 
		\midrule
		WXC & 87.59	& 117.44 & 131.90	& 151.45 & 163.48   \\
		SXC+WXC & 50.13	& 75.13	 & 78.98	& 84.17	 & 86.13   \\
		SXC & 4.92	& 9.80	 & 15.09	& 21.21	 & 27.10   \\
		\bottomrule
	\end{tabular}
\end{table}

Finally, Table~\ref{tab-time-nsf} lists the average running times (in seconds) of the heuristic algorithm for the realistic NSF network, from which it can again be observed that the results are similar to those in Table~\ref{tab-time-n6}. The heuristic algorithm can yield reasonable solutions within an acceptable running time. Thus, we can see that the proposed heuristic algorithm is also efficient for solving realistic large-scale problem instances.

\section{Conclusion}
In this paper, we focused on the resource allocation problem in SCNs, which we defined as the routing, spatial channel, and spectrum assignment (RSCSA) problem. First, we reviewed the key features of SCNs from the networking perspective and described how these features are related to the RSCSA problem. We proved the NP-hardness of the RSCSA problem and proposed two approaches for solving it: an ILP model for small-scale problem instances and a heuristic algorithm with higher scalability. Simulation results show that the ILP model (with a 1-hour running time limit) and the heuristic algorithm both work well for small-scale problem instances, for which the optimal solutions or solutions close to the lower bounds can be obtained. In addition, the heuristic algorithm is also efficient for solving realistic large-scale problem instances. Moreover, the results show that compared to conventional OXCs with full wavelength switching implemented by means of stacked WXCs, which are typically used in SDM-OTNs, the CSS-based HOXCs designed for SCNs can enable great cost savings while providing similar network performance, and consequently, these HOXCs are expected to be a promising solution for the future massive SDM era. However, some important challenges remain that have not been addressed in this paper, such as the resource allocation problem for an SCN with SLC support implemented by means of MS-based HOXCs and the dynamic resource allocation problem, which will require further investigation in future work.

\end{document}